# Long-term memory stabilized by noise-induced rehearsal

Yi Wei and Alexei Koulakov
Cold Spring Harbor Laboratory, Cold Spring Harbor, NY 11724, USA

**Abstract**

Cortical networks can maintain memories for decades despite the short lifetime of synaptic strength. Can a neural network store long-lasting memories in unstable synapses? Here, we study the effects of random noise on the stability of memory stored in synapses of an attractor neural network. The model includes ongoing spike timing dependent plasticity (STDP). We show that certain classes of STDP rules can lead to the stabilization of memory patterns stored in the network. The stabilization results from rehearsals induced by noise. We show that unstructured neural noise, after passing through the recurrent network weights, carries the imprint of all memory patterns in temporal correlations. Under certain conditions, STDP combined with these correlations, can lead to reinforcement of all existing patterns, even those that are never explicitly visited. Thus, unstructured neural noise can stabilize the existing structure of synaptic connectivity. Our findings may provide the functional reason for highly irregular spiking displayed by cortical neurons and provide justification for models of system memory consolidation. Therefore, we propose that irregular neural activity is the feature that helps cortical networks maintain stable connections.

## Introduction

Changing synaptic strengths is widely regarded as the mechanism by which long-term memory is encoded and stored in the brain (Martin et al., 2000). Long-term potentiation (LTP) and long-term depression (LTD) of synaptic conductances exhibit many features that make them obvious candidates for the cellular mechanism of memory storage. By correlating pre and postsynaptic activities, LTP/D can implement Hebbian plasticity that is at the basis of many learning and memory models (Abbott and Nelson, 2000). The persistence of synaptic changes, induced by LTP/D, could contribute to the persistence of memory traces. Because LTP is observed in many preparations, including freely behaving animals (Whitlock et al., 2006), and in many brain regions (Cooke and Bliss, 2006), it fits any reasonable description of the basic mechanism of learning and memory.

Some insight into the memory mechanism may come from comparing the lifetime of memory and persistence of synaptic changes. Although memories can be stored by the brain for dozens of years, the lifetime of LTP appears to be shorter. In hippocampal slice preparations and in vivo, the persistence of synaptic changes, in most cases, is limited by 4-5 weeks (Abraham, 2003; Shors and Matzel, 1997b). In rare instances, synaptic changes can last for about a year; however, these examples require special conditions (Abraham et al., 2002). This occurs despite the observation that at least some components of consolidated long-term memory can be attributed to the hippocampal complex (Nadel and Moscovitch, 1997, 2001). In the cortex, LTP has not been demonstrated to last beyond a period of several weeks (Ivanco and Racine, 2000; Trepel and Racine, 1998). While changes in structural connectivity can persist for more than a month (Alvarez and Sabatini, 2007; Fu and Zuo, 2011; Grutzendler et al., 2002; Knott et al., 2006; Trachtenberg et al., 2002), they may reflect ongoing changes in sensory inputs rather than carry memory traces. The same applies to other examples of cortical plasticity observed after sensory deprivation (Feldman, 2009). Whether synaptic strengths can persist throughout the lifetime is an open question (Shors and Matzel, 1997a). Cascade synaptic models (Fusi et al., 2005), for example, propose that individual synapses contain long-lasting internal states that are not directly related to synaptic strength. Another explanation is that robust long-term memories can somehow be maintained for decades without the requirement of stable synapses. This hypothesis is investigated here.

A related question, known in computational literature as plasticity-stability dilemma, poses that a memory system must evolve to be able to both store new memories promptly and retain old information (Abraham and Robins, 2005; Grossberg, 1987). Some solutions to plasticity-stability dilemma have been proposed (Carpenter and Grossberg, 1987a, 1987b). Here, we argue that, even without the challenge from novel memories, the known lifetime of LTP is not in agreement with the long-lasting nature of long-term memory. Therefore, we pose the following question: can short-lived synapses provide the basis for memories that last a lifetime?

A related question has been addressed in attempt to build molecular models of LTP (Crick, 1984). Although LTP lifetime, in most cases, is measured in weeks (Abraham, 2003), it is believed that molecules in synapses undergo turnover every several days (Lisman and Hell, 2008). Therefore, the persistence of LTP results from molecules that have relatively short lifetimes. Several studies have proposed how short-lived molecules can build a lasting synapse, including bistability (Lisman and Zhabotinsky, 2001; Miller et al., 2005) and self-sustaining molecular clusters (Shouval, 2005). This problem has many parallels with the question studied here because, in these models, relatively stable synapses result from activities of unstable molecules.

An alternative class of mechanisms involves rehearsals whereby old memories are constantly revisited and relearned via an ongoing process (Wittenberg et al., 2002). Because all old memory states must be explicitly visited within the time window determined by LTP decay, presumably in the sleep, it is unclear whether such a mechanism is realistic, especially if the number of patterns is large.

Here, we propose a mechanism for persistent memory storage that uses short-lived synapses. In our model, long-term memory can be stored in the network for a very long time, despite a short time-constant of LTP persistence. To be preserved, memory states are not revisited as in the model involving explicit rehearsals (Wittenberg et al., 2002). We analyze a simple mathematical model for attractor neural network, which includes several realistic elements, such as stochastic neural noise, short synaptic lifetime, and ongoing synaptic plasticity described by spike-timing dependent plasticity (STDP). The network resides near a set of states that represent activity relevant to its current environment. The average activity of neurons samples only these current memory states. However, we demonstrate that, because of the presence of noise, the correlations in neural activity carry imprints of all memory traces, including old ones. These correlations, under carefully chosen conditions, can allow the old traces to be rehearsed and maintained by the network, even though they are not explicitly visited. We thus propose that old memory states can be reinforced by rehearsal, even though these memories are never visited or accessed. Because our rehearsal mechanism does not involve explicit reactivation of old memories, we call our process implicit rehearsal. We show that, for implicit rehearsal to be effective in reinforcing old memory states, STDP rules must satisfy certain strict conditions. This mechanism will work with antisymmetric STDP that is often observed (Bi and Poo, 2001; Froemke and Dan, 2002) but does not work with the symmetric non-negative form of LTP. We show that neural noise combined with synaptic plasticity can lead to stability of old memory traces, despite individual synapses being unstable. Our model has experimentally testable predictions.

## Results

**Patterns of neural activity stored in network weights correspond to network attractors.** In this paper, we analyze attractor neural networks with features similar to the continuous Hopfield model (Hertz et al., 1991a; Hopfield, 1984). Such networks can exhibit two types of memory: long-term, contained in the recurrent network weights, and working memory, contained in the firing rates of neurons (Amit, 1989).

The network can store the long-term memory of a set of patterns in pre-specified network weights (Hertz et al., 1991a). This is intended to model the long-term memory that is pertinent to neuronal networks. If activity of a neuron number $i$ that is associated with pattern number $a$ is $P_i^a$, then, as within the conventional Hopfield model, the connection strength between two neurons $i$ and $j$ is given by the Hebbian-like learning rule:

$$W_{ij}(t) = \frac{1}{N}\sum_{a=1}^{p} c_a P_i^a P_j^a . \qquad (1)$$

This means that in the more patterns a given pair of neurons is coactive, the stronger the connection between these neurons. Here, $N$ and $p$ denote the total number of neurons and the number of stored patterns, respectively. We also introduced a set of coefficients, $c_a$, that describe how strongly a given pattern is included in the network connections. In the standard Hopfield model, these coefficients are initialized and remain equal to one. In this paper, these coefficients are affected by the ongoing activity in the network. The goal of our study is to understand the long-term behavior of the strengths of the patterns $c_a(t)$ that result from ongoing learning.

In this network, patterns that are embedded in the recurrent weights, according to equation (1), become network attractors (Hopfield, 1982, 1984). This means that if the activity of neurons matches one of the patterns at some moment in time, the pattern will be maintained by recurrent connections, despite small perturbations and noise. Because several patterns are simultaneously embedded in the weights, the network may have several stable attractor states, provided that the number of patterns, $p$, is not too large (Hertz et al., 1991b). Because network firing rates can persist only near an attractor, in the absence of external inputs, the network must choose where to reside. This decision can be viewed as an implementation of short-term memory of the "which attractor I am near" kind. Therefore, Hopfield nets can support both short (working) and long-term memory (Bird and Burgess, 2008; Cowan, 2008) (Figure 1).

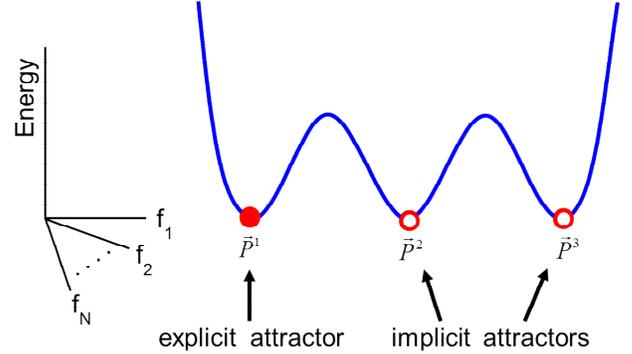

**Figure 1.** Attractor neural networks. Dynamics of neuronal firing rates are viewed as gradient descents with the landscape defined by an energy-like function (blue) (Hopfield, 1982, 1984). The landscape is defined in the space of activities of all neurons in the network ($f_1…f_N$). Position in the landscape is determined by the combined vector of neural activities. At the bottom of the landscape are the patterns that are stored in network weights according to equation (1). These patterns represent network attractors that encode long-term memory stored in the network. In our study, attractors are divided into two classes: explicit and implicit. Explicit states (full circle) are actively explored by the network, therefore, can be reinforced by rehearsal. Implicit attractors (empty circles) have not been visited by the network within the time window of synaptic decay. Therefore, these patterns are expected to disappear because of the decay of synaptic strengths.

**STDP rules applied to the attractor neural network lead to the deterioration of stored memories.** What is the effect of synaptic plasticity on the attractors that are embedded into the network? From the point of view of plasticity, it is important to distinguish two types of attractors. First, there are attractors that represent memories that the network is constantly visiting. For example, because of external stimuli, the network can hop around states that are relevant to the particular task or environment. These attractor states represent recent memory. More precisely, recent states are defined as those that are visited within the time constant of synaptic decay. We call this type of states explicit attractors. The other type of state represents memories that were embedded into the network a long time ago and have not been accessed recently. These states will be called implicit (Figure 1). Please note that our terminology is somewhat different from the convention that uses the terms explicit/implicit to denote different classes of memory (i.e., declarative and procedural) (Schacter, 1987).

Synaptic learning leads to different outcomes for explicit and implicit memory states. Because explicit memories are replayed in network activities, they are constantly rehearsed. Thus, their contribution within the weight matrix is stable. In the Methods section, we evaluate the component of the weight matrix that carries explicit states [equation (14)]. Specifically, we show that this component does not decay with time and is reinforced by learning in the network.

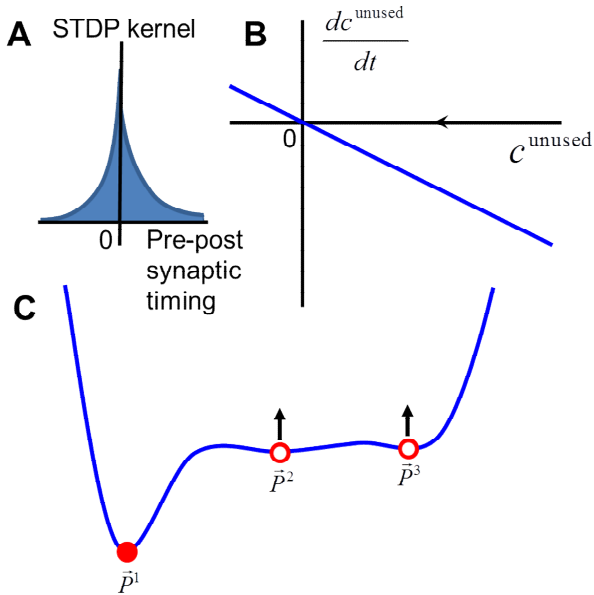

**Figure 2.** Synaptic plasticity implemented in the attractor neural network destroys implicit states. (A) STDP rules used. The rate of change of connection strength between two neurons (vertical axis) as a function of differences in spike time between pre and postsynaptic cells. (B) The rate of change in the coefficient with which an implicit pattern enters the synaptic weight matrix, defined by equation (1), as a function of the value of the coefficient. This coefficient describes the strength of the pattern in the weight matrix. For positive values of the coefficient, the rate of change is negative (left arrow), which implies decay. The decay is exponential $c^{unused} \sim \exp(-t/\tau_0)$, where $\tau_0$ is the time-constant of synaptic decay. (C) Illustration of the implications of decay of the coefficient for network attractors. Implicit attractors become less stable and disappear because they vanish from the weight matrix.

The behavior of implicit memories is quite different. If the attractors that correspond to implicit patterns are not visited within the time window of the decay of synaptic strength, defined in our model by parameter $\tau_0$, these memory states disappear from the network weight matrix (Figure 2). This observation is not surprising because rehearsal that reinforces the explicit attractors is not available for implicit attractor states. This is because the latter are not present in the network activity.

**Noise added to the network can implement rehearsal of old (implicit) memory states.** Next, we included noise in the inputs of neurons to test whether noise can reinforce implicit memory states. We reasoned that if white unstructured noise were added to the input of every neuron, the activity of the network would contain implicit memory states, which may potentially stabilize old memories through the process of rehearsal. We call the process of rehearsal that is based on random noise implicit rehearsal. This process is distinct from the rehearsal of explicit states that occurs due to the network actually visiting explicit attractors.

The dynamics of implicit rehearsal are as follows. Random unstructured noise is added to the inputs of every neuron in the network. The term unstructured implies that the amount of noise added to neurons does not contain the patterns being rehearsed. In this study it is assumed to be the same for all neurons for simplicity. Because neurons are connected by recurrent weights that do contain implicit patterns, when noise passes through recurrent connections, it becomes structured. This means that neural activity acquires correlations that contain implicit patterns [equation (18)]. This is because implicit states are amplified by positive feedback that is present within the recurrent weight matrix [equation (1)] and fluctuations along these directions are therefore amplified by recurrent connections. Thus, despite the network staying near explicit attractors and never visiting the implicit states, the presence of implicit states in the weight matrix shapes network fluctuations along the directions that represent old memories. Implicit memory is contained in the correlations of network activity as opposed to the explicit memory that is contained in the mean firing rate.

**Non-negative symmetric STDP rules applied to the network with white noise do not stabilize old (implicit) memory.** What is the effect of implicit rehearsal in the case when STPD rules are ongoing in the network? For an STDP learning rule, a change in synaptic efficacy is dependent on the relative timing of pre and postsynaptic action potentials for every synapse. In the simplest case, the synapse becomes stronger if both presynaptic spikes precede the action potential in the postsynaptic neuron and, in the opposite case, when postsynaptic spikes precede presynaptic spikes. We call this form of learning rule symmetric non-negative (Figure 3). Our results show that this type of learning rule, applied in the presence of neural noise, does not make unused (implicit) memory more stable. The contribution of a memory state into the weight matrix is determined by coefficient, $c$, defined by equation (1). The rate of change in this contribution defines the behavior of old memories with time. Figure 3B shows that, in the case of non-negative

symmetric STDP, the only stable point for this coefficient is zero, which means that implicit memory is destined to disappear [see equation (19) for more detail]. Interestingly, if the strength of this coefficient is sufficiently large and if it passes the transition point in Figure 3B, the coefficient becomes unstable. This instability implies that the old pattern will emerge spontaneously in the network when the strength of the pattern is sufficiently large. In both regimes of small and unstable $c$, the network cannot maintain the old memory in a reliable manner.

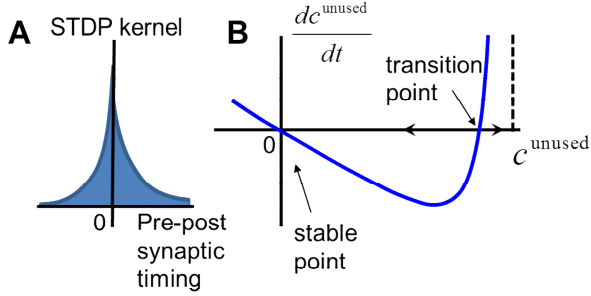

**Figure 3.** Unstructured neural noise does not stabilize implicit attractors in the case of the non-negative symmetric STDP rule. (A) Synaptic learning rule is the same in Figure 2A. (B) The rate of change in the contribution of the implicit attractor to the weight matrix contains only one stable point at $c^{unused} = 0$, which implies a lack of stability of unused memory. This dependence is represented by equation (19).

**Antisymmetric STDP rules combined with white noise can stabilize old memory states that are not revisited.** We examined the stability of implicit states when STPD rules have a form that is often observed experimentally (i.e., antisymmetric) (Bi and Poo, 2001; Froemke and Dan, 2002; Sjostrom et al., 2008). We assumed that if a presynaptic spike precedes the postsynaptic spike, the synapse is strengthened due to LTP. If the timing of the spikes is reversed, the synapse is weakened (i.e., the contribution of such events to the synaptic strength are negative (Figure 4A)), which corresponds to LTD. We find that, in this case, an implicit (unused) memory state can have two stable points. The stable points are defined for the contribution of the pattern to the weight matrix $c$ [equation (1)]. Stable points can be determined by examining the rate of change of this contribution (Figure 4B) that is given by equation (19). If, for a certain value of contribution, the rate of change is zero, this value is called the stationary point. If the contribution of a pattern is placed exactly into one of the stationary points, it will remain there because the rate of change of $c$ is zero. Figure 4B shows three stationary points in this case. These three points differ in cases of small perturbations that deflect contribution, $c$, slightly from a stationary state. For two of the stationary states in Figure 4B, the resulting rate of change returns the contribution back to the state. This is illustrated by the arrows on the horizontal axis. Therefore, these two states are stable. The third stationary point is unstable and is called the transition point.

For two stable states, the contribution of the pattern is either low (stable point 1) or high (stable point 2). The former state corresponds to the weak representation of the pattern in the network that is indistinguishable from noise. The high contribution point (stable point 2) corresponds to the memory that is substantively present in the weight matrix. The system is capable of maintaining either high or low levels of a pattern in the weight matrix virtually indefinitely. This is despite the decay of synaptic strength that is ongoing in the system. Contribution is maintained because noise implements implicit rehearsal. Although the average values of firing rates are near the explicit attractor, the correlations in the firings rates between cells, induced by noise, carry information about other patterns that are not visited (18). Because learning rules are dependent on correlations, Hebbian learning is capable of maintaining implicit states in memory. This correlation-induced rehearsal results in the stability of patterns as a function of time. Although we presented the results for a single pattern, other implicit states are stabilized similarly due to their independence [equation (3)]. Implicit rehearsal can stabilize several patterns simultaneously.

**Conditions of bistability.** For the pattern contribution to have two stable points, several conditions must be met. First, the integral of STDP kernel (Figure 4A) must be negative. This implies that the LTD part of the STDP curve is stronger than the LTP part. Second, we need $\gamma g^2 \xi^2 (A_+ + A_-) > \frac{2\tau^2}{\tau_+ \tau_-}$, which can be satisfied if time scales of STDP, $\tau_\pm$, are larger than that of firing rate, $\tau$. This condition guaranties that $\frac{dc_a}{dt}$, given by equation (18), has one local minimum and one local maximum on the region $0 \leq c_a(t) < 1/g$. To have the second stable point at finite $c_a(t)$, we also need the local maximum to be positive. In the case that $\gamma g^2 \xi^2 (A_+ + A_-)$ is much larger than $\frac{2\tau^2}{\tau_+ \tau_-}$, this condition can be met when $\frac{\gamma g^2 \xi^2}{2} \left( \frac{A_+ + A_-}{3} \right)^3 > \left( \frac{\tau}{2\tau_+ \tau_-} (A_+ \tau_+ + A_- \tau_-) \right)^2$. Figure 4B shows the typical behavior of $\frac{dc_a}{dt}$ and a set of values for parameters.

**Our model predicts correlations between network weights and neural noise.** What are the implications of our findings for an individual synapse? To maintain a set of memories in the network, the synapses have to preserve their strength. Because, in our model, synaptic strength decays with time, it has to be maintained at a constant level by the correlations in the pre and postsynaptic activities. This means that stronger synapses have higher correlations in pre and postsynaptic activities. Thus, the form of rehearsal proposed here can be detected by measuring the correlations in activity for individual synapses and observing their relationships with synaptic strength. A more precise definition of this relationship is given by equation (11).

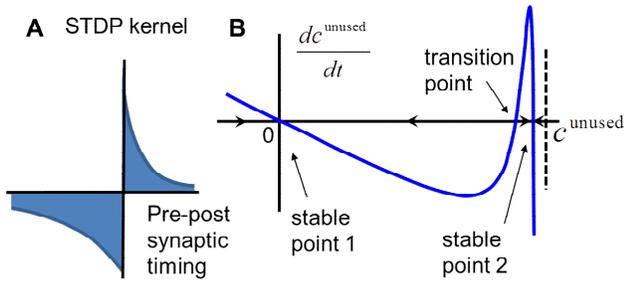

**Figure 4.** The stabilization of old memory states by the combination of unstructured noise and antisymmetric STDP learning rule. (A) Antisymmetric STDP learning rule. (B) The rate of change of the contribution of an unused state ( $dc^{unused}/dt$ ) as a function of the contribution itself ( $c^{unused}$ ). This dependence is represented by equation (19). In this case, the contribution has two stable points, near zero and at a finite value. The former/latter stable points correspond to the unused memory pattern being absent/present in the network connectivity, respectively. At a stable point, the rate of change of the pattern contribution is zero. In addition, small perturbations from the stable point will induce the rate of change that returns the system back. At the transition point, the rate of change is zero and unstable. Parameters used are $A_+^{'} = 0.02$ , $A_-^{'} = -0.012$ , $\tau_+ = 50\text{ms}$ , $\tau_- = 100\text{ms}$ , and $\tau = 5\text{ms}$ .

## Discussion

In this paper, we studied the behavior of memory states stored in the weights of an attractor neural network. Our network included some realistic features, such as STDP learning rules, neural noise, and limited LTP/D lifetime. We assumed that leaning occurs in the network on a continuous basis (i.e., the weights are continuously updated to reflect ongoing activity). In these conditions, the network weights should reflect the ongoing activity that we described by the term explicit attractors. The other set of states, which we called implicit attractors, represent memories that were stored at some point in the past and that have not been revisited recently or within the time-constant of synaptic decay. Such states are expected to disappear from the network weight matrix because of the decay of synaptic strengths. How can the network maintain implicit memory states despite synaptic decay?

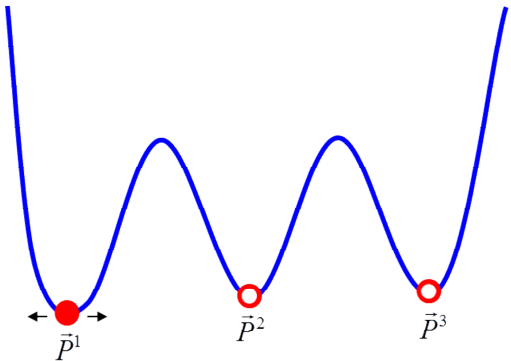

**Figure 5.** Illustration of the stability of implicit (unused) states. Implicit attractors 2 and 3 do not vanish with time, rather remain stable due to the fluctuations (arrows) around explicit attractor 1. Although the average activity remains near the explicit attractor, the fluctuations are biased toward the direction of implicit states, which leads to their rehearsal.

We show that unstructured noise can substantially alter the dynamics of learning. Although input noise is unstructured in our model (i.e., it does not contain stored patterns), when noise passes through recurrent synaptic weight matrix, it becomes colored. This means that the correlations in neural activity that are induced by noise reflect all memory patterns (explicit and implicit) that are stored in the network. This is because the memory states represent the directions in neural activity that are amplified by the positive feedback present in the recurrent network. Thus, although the average neural activity represents recent states only, the correlations reflect the entirety of memories, including the old ones (i.e., implicit). Because synaptic learning is dependent on correlations, in principle, it can reinforce implicit memory states, when certain conditions are met. We show that the antisymmetric STDP rule, which contains both positive and negative components (i.e. both LTP and LTD (Bi and Poo, 2001; Froemke and Dan, 2002), can reinforce old memory traces without explicitly visiting them (Figure 4). In contrast, non-negative symmetric STDP cannot stabilize old memories (Figure 3).

In our model, the old memory traces (implicit) are never visited or accessed by the network. The network always resides near the set of newer states that are relevant to current behavior and, therefore, are called explicit. Yet, we propose, that implicit states can be rehearsed. The rehearsal occurs not because the average activity but fluctuations reflect the old states. We call this form of rehearsal in which the old memory is never directly accessed, implicit rehearsal.

Our model suggests the reason for the high degree of irregularity observed in firing of cortical neurons (Softky and Koch, 1993). In our model, irregular neuronal firing implements rehearsal of old memory patterns. When neural noise is passed through the network weight matrix, it captures the information about patterns stored in these connections. The ongoing synaptic plasticity can subsequently reinforce the stored patterns. Neural noise plays an essential role in generating correlations in neural activity. Similar roles can be played by the unreliable nature of synaptic vesicle release (Sudhof, 2004). Thus, we propose that unreliable neural activity is the feature that helps cortical networks maintain stable connections.

In our model, both network weights and firing rates exhibit attractor behaviors. Thus, firing rates can have several discrete states that are robust with respect to small perturbations and noise and are called network attractors. The identity of these states depends on the strengths of recurrent connections between neurons (Amit, 1989; Amit et al., 1994). Additionally, these states represent long-term memories that are stored in the network weight matrix. In our model, network weights also exhibit attractor behaviors. We show that, because of ongoing synaptic plasticity, a weight matrix can have self-maintaining stable states that could also be called attractors. Thus, in Figure 4B, we show that synaptic weight matrix can have two stable states that correspond to a given memory pattern being present or absent from the network connectivity. Once the weight matrix is placed near the state that includes a given memory pattern, it will stay there for a long time, which explains the stability of the memory of the pattern. The attractors of the weight matrix, in our model, are maintained by ongoing neural activity generated by network noise. As such, we argue that neural activity helps synaptic weights form stable states (i.e., attractors). In our model, firing rates and synaptic connections form two dual systems of attractors: synaptic weights help firing rates to exhibit discrete stable states while firing rates stabilize discrete self-maintaining states within network connections.

Our model can provide a rationale to the standard model of system consolidation (Dudai, 2004). We posit that certain memory traces can be maintained in stable states over long periods by implicit rehearsal. The problem of placing the network into these states is not addressed here. However, we notice that the regions of stability surrounding stable memory states are narrow (stable point 2 in Figure 4B). The parameters of network weights that describe the contribution of a given memory pattern must be tuned to a relatively precise value. We argue that the function of placing the network in a narrow parameter range, which is necessary for long-term storage, is performed by memory consolidation. Once consolidated, (i.e., placed near stable point 2 (Figure 4B)), a memory pattern is maintained by implicit rehearsal. Thus, we argue that the functional role of system memory consolidation is to place the network weights within the narrow range of parameters where the memory trace can persist for a long time.

Our model provides experimentally testable predictions. Within the implicit rehearsal mechanism, synaptic strength is larger for synapses with stronger correlations between pre-and postsynaptic activities [equation (11)]. The remainder of the network produces these correlations, which then reinforce the synaptic strength. This prediction can be tested if synaptic strength is measured simultaneously with ongoing neural activity for individual synapses. In doing so, one should isolate correlations of activity induced by measured synapse and the remainder of the network. This could be done pharmacologically or by including correlations over certain time-scales, such as one temporal semiaxis for unidirectional synapses. Specific STDP kernel could also be surmised based on studies of synaptic plasticity or could be derived from the best match between synaptic strength and activity correlations. Overall, we propose that synaptic strengths are maintained by ongoing irregular spiking, which can be tested experimentally.

An interesting observation is that stability of long-term memory in our model seems to imply stability of individual synapses, which is in contrast to the fleeting nature of synaptic strengths discussed in the introduction. Although our model does stabilize memory states, it also allows individual synapses to be unstable. Because, in our model, memory is delocalized and each memory trace is represented by nearly all synapses in the network, variability in individual synapses does not imply memory decay. Our model offers two distinct scenarios for how memory states are corrupted. First, they can completely disappear through a discontinuous jump between two stable points, as in Figure 4B. Second, they can slowly change by changing each individual synapse at a time, whereby a memory state morphs into something else. Because the number of synapses involved in each trace is large, this process may progress for a long time without substantial change in the representation of memory. Overall, our model predicts that synaptic lifetime observed in reduced preparations, such as in slices, should be shorter than in vivo, because the latter lifetime is improved by the ongoing activity. Rare instances, when synapses show stability (Abraham et al., 2002), could be attributed to the mechanism of stabilization proposed here. In vivo, synaptic persistence can be shorter than memory retention time because memory is a collective property of a large ensemble of synapses.

**Conclusion**

We study the stability of long-term memory patterns stored in a recurrent neural network. Our model includes ongoing synaptic plasticity regulated by STDP rules. We show that old memory traces can be stabilized by fluctuations of neural activity when STDP rules satisfy certain constraints. Old memory patterns become stable self-maintaining and persistent states of the network weight matrix. Our model provides a mechanism for the extension of memory lifetime via the combination of ongoing synaptic plasticity and neural noise.

## Materials and methods

**Description of the model.** Let $N$ be the total number of neurons. We will assume that there are $p$ patterns represented by $N$-dimensional vectors, whose elements are ±1:

$$P_i^a = \pm 1, \quad a = 1,\ldots,p, \quad i = 1,\ldots,N. \quad (2)$$

Different patterns are chosen to be orthogonal to each other,

$$\frac{1}{N} P^{aT} \cdot P^b = \delta^{ab}. \quad (3)$$

Equation (3) allows us to define $p$ projection operators as

$$\hat{P}^a_{ij} = \frac{1}{N} P_i^a P_j^a, \quad \text{so that} \quad \hat{P}^a \cdot \hat{P}^b = \delta^{ab} \hat{P}^a. \quad (4)$$

When projection operators number $a$ are applied to a given activity vector, they result in an activity specific to the given pattern $a$.

Memories about the patterns are stored in a synaptic weight matrix using the conventional learning rule (Dayan and Abbott, 2001),

$$W_{ij}(t) = \sum_{a=1}^{p} c_a(t) \hat{P}^a_{ij}, \quad (5)$$

where the set of coefficients, $c_a(t)$, represents the strengths of individual patterns.

In our model, the input current and firing rate of neurons $i$, $u_i(t)$, and $f_i(t)$, are related through the activation function $F$

$$f_i(t) = F(u_i). \quad (6)$$

Input currents are described by the equation

$$\tau \frac{d\vec{u}(t)}{dt} = -\vec{u}(t) + \hat{W} \vec{f}(t) + \vec{\xi}(t). \quad (7)$$

In equation (7), $\tau$ is a constant, which determines how rapidly the current varies and $\vec{\xi}(t)$ is the Gaussian random white noise. We assume that $\vec{\xi}(t)$ has the following properties

$$\langle \xi_i(t) \rangle = 0, \quad \langle \xi_i(t)\xi_j(t') \rangle = \xi^2 \delta_{ij} \delta(t-t'), \quad (8)$$

where $\langle \cdots \rangle$ denotes the average over noise ensemble. Subsequently, we assume that the amplitude of noise, $\xi$, is very small and we use this fact along with the short time scale of $\xi$ to treat noise as a perturbation.

Plasticity in the network is defined by spike timing dependent learning rules. For a pair of cells, the strength of synapses is updated with a rate of update that is dependent on pre and postsynaptic activities:

$$\tau_0 \frac{dW_{ij}}{dt} = -W_{ij} + \gamma \int_{-\infty}^{t} dt_1 \langle f_i(t_1) K(t_1 - t) f_j(t) \rangle + \gamma \int_{-\infty}^{t} dt_2 \langle f_i(t) K(t - t_2) f_j(t_2) \rangle. \quad (9)$$

Here, $f_i(t)$ is the firing rate of neuron number, $i$, at time $t$, and $\gamma$ is the learning rate. Three terms in the r.h.s. of this equation describe the decay of synaptic strength with time; the modification due to presynaptic and postsynaptic firing, respectively. The STDP kernel, $K(\Delta t)$, in our model contains two components: short-range and long-range, $K(\Delta t) = K_s(\Delta t) + K_l(\Delta t)$. We assume that the short-range component varies within the time scale of several hundred milliseconds, which is defined as:

$$K_s(\Delta t) = \begin{cases} A_+ \exp(\Delta t / \tau_+) & \Delta t < 0 \\ A_- \exp(-\Delta t / \tau_-) & \Delta t \geq 0. \end{cases} \quad (10)$$

The long-range STDP kernel, $K_l(\Delta t)$, is needed in our model to constrain the overall magnitude of firing rates and can originate from metabolic and other constraints. We assume that it varies very slowly on the time scales of the order of hours or more. The only constraint on $K_l(\Delta t)$ that is important in our model is that it makes the integral of the entire STDP kernel over time positive (i.e., $\int_{-\infty}^{\infty} K(t) dt > 0$) as discussed in more detail following equation (14).

STDP rule (9) can also be rewritten in an equivalent integral form

$$W_{ij}(t) = \frac{\gamma}{\tau_0} \int_{-\infty}^{t} dt_1 \int_{-\infty}^{t} dt_2 \, e^{-\frac{t - \max(t_1, t_2)}{\tau_0}} \langle f_i(t_1) K(t_1 - t_2) f_j(t_2) \rangle. \quad (11)$$

This equation shows that $\tau_0$ defines the forgetting time-constant. The old memory is expected to decay after this time with the exception of memory that is rehearsed (i.e., relearned within the time scale $\tau_0$). This rehearsal process is the topic of our present study.

Due to random noise, $\vec{\xi}(t)$, the input currents fluctuate near constant values $\vec{u}(t) = \vec{u} + \delta\vec{u}(t)$. We assume that fluctuations are weak and can be treated as small perturbations [see discussion following equation (16) for the justification of this assumption]. Therefore, by equation (6), the firing rates fluctuate around stationary rates

$$\vec{f}(t) \approx \vec{f} + g\delta\vec{u}(t), \qquad (12)$$

where $g = F'(u_i)$. Because neural noise is short-range, its time-scales are measured in milliseconds and we can decompose the dynamics of the system into two components: the fast-changing component, associated with noise, and the slowly varying component, determined by Hebbian learning. These two components are represented by two terms in equation (12). The equation for fast-changing component is

$$\tau \frac{d\delta\vec{u}(t)}{dt} = -\delta\vec{u}(t) + g\hat{W}\delta\vec{u}(t) + \vec{\xi}(t). \qquad (13)$$

Our goal is to derive the contribution of the fast-changing component (i.e., noise) to the slowly-varying component. This interplay could be interpreted as rehearsal. In the subsequent discussion, we treat noise as a small perturbation to the firing rates of the network (i.e., we will assume that $\xi^2 \ll 1$).

For the given set of network weights, we assume that there are two types of attractors. One set of attractors is never visited by the network. We call these states implicit. The other set of attractors is explicitly visited by the system. Because our main goal is to consider the dynamics of implicit attractors (i.e., ones that are never visited), for simplicity, we assume that the explicit attractors are represented by only one attractor. Here, we discuss briefly the explicit attractor state and its stability with respect to learning.

Let us assume that the explicit attractor has an index $a = 1$. The stationary firing rates associated with this state are proportional to the pattern $P^1$, i.e. $\vec{f} = bP^1$.

Here, $b$ is a constant determined by the function $F$ (we assume that $F(x) = -F(-x)$, e.g., $F$ is the sigmoid function). Assuming that the effects of noise are negligible, one can obtain the stationary value of the weight matrix, which results from the explicit attractor. This contribution is present by virtue of the explicit attractor relearning, itself, through the STDP rules and could be viewed as resulting from explicit rehearsal (i.e., rehearsal of patterns that are currently in working memory). For this type of rehearsal, noise is not necessary. Equation (11) allows us to determine this component of the weight matrix

$$W_{ij} = c_1 \hat{P}^1_{ij} + \delta W,$$
$$c_1 = \gamma b^2 (A_+ \tau_+ + A_- \tau_- + \Delta). \qquad (14)$$

where $\delta W = \sum_{a \neq 1} c_a \hat{P}^a_{ij}$ is the component of the weight matrix as determined by other (implicit) patterns. From equations (6) and (14), factor $b$ is determined by the self-consistent equation,

$$F\left[\gamma b^3 (A_+ \tau_+ + A_- \tau_- + \Delta)\right] = b. \qquad (15)$$

Note, that we assume that $|F''(u_i)|$ is very small. This component, $\delta W \to 0$, is in the network without noise according to equation (11). Our goal here is to determine the behavior of $\delta W$ (i.e., the contribution of implicit patterns that are never explicitly visited) in the network with noise.

The parameter $\Delta \equiv \int_{-\infty}^{\infty} K_l(t)dt$ makes the integral of STDP kernel positive so that $\int_{-\infty}^{\infty} K(t)dt = A_+\tau_+ + A_-\tau_- + \Delta > 0$. Because this is the only point at which the long-range STDP kernel $K_l(t)$ enters our model, we will not discuss this kernel further. From this point on, by STDP kernel we imply the short-range kernel, $K_s(t)$, that varies on the time scales of hundreds of milliseconds and is usually measured in LTP/D experiments (Abbott and Nelson, 2000).

According to equation (11), without noise, the weight matrix contains only pattern $P^1$ with stationary strength $c_1$. All other coefficients $c_{a \neq 1}$ are zero. With Gaussian white noise included in equation (7), other patterns (implicit) are represented in the activity of the network. Therefore, these patterns are present in the synaptic weight matrix $\delta W(t)$. Our goal is to find how coefficients $c_{a\neq 1}(t)$ evolve with time in this case.

To accomplish this goal, we consider noise a small perturbation and use the perturbation theory using the amplitude of noise, $\xi^2$, as a small parameter. Noise-induced firing rate fluctuation, $\vec{\delta u}(t)$, in the direction of pattern number, $a$, is $\vec{\delta u}^a(t) = \hat{P}^a \cdot \vec{\delta u}(t)$ and the corresponding component of random inputs is $\vec{\xi}^a(t) = \hat{P}^a \cdot \vec{\xi}(t)$. By applying projector, $\hat{P}^a$, on both sides of equation (13), and because $c_a$ changes much slower than $\vec{\delta u}(t)$, we find

$$\vec{\delta u}^a(t) = \frac{1}{\tau}\int_{-\infty}^{t} dt' e^{-\frac{t-t'}{\tau}(1-gc_a)} \vec{\xi}^a(t') . \quad (16)$$

Next, we derive the condition for the validity of perturbation theory (i.e., the amplitude of fluctuations due to noise is smaller than the zero-th order solution obtained without noise). Using equations (18) and (19), we can estimate $\langle \delta u_i^2 \rangle \propto \xi^2 \cdot \frac{\tau_\pm}{\tau}$. $u_i = bc_1$, where $c_1$ and $b$ are given in equation (14-15). By choosing $\xi$ to be small, we can make $\delta u_i$ much smaller than $u_i$ to allow us to treat noise as a perturbation in equation (12).

By equations (3) and (7), we have

$$\langle \vec{\xi}^a(t_1)\vec{\xi}^b(t_2) \rangle = \xi^2 \hat{P}_{ij}^a \delta^{ab} \delta(t_1 - t_2) . \quad (17)$$

Using equations (8) and (16), we obtain the average correlation function of fluctuations

$$\langle \delta u_i^a(t_1) \delta u_j^b(t_2) \rangle = \frac{\xi^2 \delta^{ab}}{2\tau N} \frac{1}{1-gc_a} e^{-\frac{|t_1-t_2|}{\tau}(1-gc_a)} P_i^a P_j^a \quad (18)$$

By placing this correlation function into equation (9), we obtain the equations that describe the dynamics of "unused" components of the weight matrix ($a = 2, \cdots, p$):

$$\tau_0 \frac{dc_a}{dt} = -c_a + \frac{1}{1-gc_a}\left(\frac{A_+'}{\frac{1-gc_a}{\tau}+\frac{1}{\tau_+}} + \frac{A_-'}{\frac{1-gc_a}{\tau}+\frac{1}{\tau_-}}\right). \quad (19)$$

Coefficients $c_a(t)$ are defined in equations (1) and (5). Here, we defined the parameters of the short range STDP kernel as

$$A_\pm' = \frac{\gamma g^2 \xi^2}{2\tau} A_\pm . \quad (20)$$

Equation (19) describes how the strength of each unused pattern's representation in the network weights changes over time when neurons receive random noise. This equation is the main result of this paper. The dependence of the right rand side of equation (19), as a function of $c_a$, is shown in Figures 2B-4B.

It is evident from equation (19) that $c_a(t) = 1/g$ is a critical value at which the equation for $c_a(t)$ becomes singular.

The equation of evolution for the first pattern is

$$\tau_0 \frac{dc_1}{dt} = -c_1 + \frac{1}{1-gc_1}\left(\frac{A_+'}{\frac{1-gc_1}{\tau}+\frac{1}{\tau_+}} + \frac{A_-'}{\frac{1-gc_1}{\tau}+\frac{1}{\tau_-}}\right),$$
$$+ \gamma bh \frac{\xi^2}{2\tau}(A_+\tau_+ + A_-\tau_- + \Delta)\sum_{a=1}^{p}\frac{1}{1-gc_a}$$

where $c_1$ satisfies the self-consistent equation $F[bc_1] = b$ and $h = F''(u_i)$ is the second derivative of the activation